\documentclass[conference]{IEEEtran}

\usepackage{cite}
\usepackage{amsmath,amssymb,amsfonts}
\usepackage{algorithmic}
\usepackage{graphicx}
\usepackage{textcomp}
\usepackage{xcolor}

\usepackage{tikz}
\newcommand*\circled[1]{\tikz[baseline=(char.base)]{
            \node[shape=circle,fill=black,draw,text=white,inner sep=1.2pt] (char) {#1};}}

\usepackage{url}
\usepackage{enumitem} %

\usepackage{upgreek} %

\usepackage{listings} %

\usepackage{tabularx} %
\usepackage{booktabs} %
\usepackage{multirow}
\usepackage{hhline} %
\usepackage{pgfplotstable} %

\usepackage{easybmat} %

\usepackage{subcaption} %
\usepackage{rotating}		%

\usepackage{xspace} %

\usepackage[ruled,noend,linesnumbered]{algorithm2e} %

\DontPrintSemicolon     %
\SetNlSty{}{}{}                %
\SetAlgoInsideSkip{smallskip}   %
\SetAlCapSkip{1.5mm}             %
\SetInd{1.5mm}{1.5mm}             %

\SetAlFnt{\small}			%
\SetAlCapFnt{\small}		%
\SetAlCapNameFnt{\small}

\newtheorem{definition}{Definition}

\newtheorem{questionW}{Question}
\newtheorem{resultW}{Result}

{\end{itshape}
}
\let\footnotesize\scriptsize
\usepackage{hyperref}   %
\hypersetup{                                                            %
	pdfpagemode=UseNone,               %
    pdfstartview=Fit,
    pdfpagelayout=SinglePage,       %
    bookmarksnumbered=true,
    bookmarksopen=false,             %
    unicode,
    colorlinks=true,       %
    linkcolor=blue,          %
    citecolor=cyan,  %
    filecolor=magenta,      %
}

\usepackage[capitalise,nameinlink]{cleveref} %

\crefname{algocf}{alg.}{algs.}
\Crefname{algocf}{Algorithm}{Algorithms}
\crefalias{AlgoLine}{line}%
\usepackage{tcolorbox}		%
\tcbuselibrary{breakable,skins}		%

\tcbset{examplestyle/.style={
		enhanced jigsaw,	%
		colback=blue!08,	%
		colframe=blue!08,	%
		arc=2mm,			%
		boxrule=0pt,		%
		left=1mm,
		right=1mm,
		left skip=0mm,  %
		right skip=0mm, %
		top=1mm,		%
		bottom=1mm,		%
		breakable,		%
		parbox = false,		%
		before={\par\pagebreak[0]\vspace{1mm}\parindent=0pt},		
		after={\par\pagebreak[0]\vspace{1mm}\parindent=0pt},				
		bottomrule = 0mm,
		boxsep = 0mm,					%
		topsep at break=0pt,			%
		bottomsep at break=0pt,			%
		pad at break=0mm,
		pad before break=1mm,		
		pad after break=1mm,		
		bottomrule at break=0mm,
		toprule at break=0mm,		
		}}

\tcolorboxenvironment{definition}{examplestyle}

\makeatletter
\DeclareRobustCommand*\uell{\mathpalette\@uell\relax}
\newcommand*\@uell[2]{
  \setbox0=\hbox{$#1\ell$}
  \setbox1=\hbox{\rotatebox{10}{$#1\ell$}}
  \dimen0=\wd0 \advance\dimen0 by -\wd1 \divide\dimen0 by 2
  \mathord{\lower 0.1ex \hbox{\kern\dimen0\unhbox1\kern\dimen0}}
}
\makeatother

\newcommand{\smallsection}[1]{\vspace{2mm}\noindent\textbf{#1.}} %

\renewcommand{\epsilon}{\varepsilon} %

\usepackage{booktabs} %
\graphicspath{{./figs/}}
\usepackage{numprint}
\usepackage{multirow}

\usepackage{scalerel}
\usepackage{tikz}
\usetikzlibrary{svg.path}

\definecolor{orcidlogocol}{HTML}{A6CE39}
\tikzset{
  orcidlogo/.pic={
    \fill[orcidlogocol] svg{M256,128c0,70.7-57.3,128-128,128C57.3,256,0,198.7,0,128C0,57.3,57.3,0,128,0C198.7,0,256,57.3,256,128z};
    \fill[white] svg{M86.3,186.2H70.9V79.1h15.4v48.4V186.2z}
                 svg{M108.9,79.1h41.6c39.6,0,57,28.3,57,53.6c0,27.5-21.5,53.6-56.8,53.6h-41.8V79.1z M124.3,172.4h24.5c34.9,0,42.9-26.5,42.9-39.7c0-21.5-13.7-39.7-43.7-39.7h-23.7V172.4z}
                 svg{M88.7,56.8c0,5.5-4.5,10.1-10.1,10.1c-5.6,0-10.1-4.6-10.1-10.1c0-5.6,4.5-10.1,10.1-10.1C84.2,46.7,88.7,51.3,88.7,56.8z};
  }
}

\RequirePackage{etoolbox}
\DeclareRobustCommand\orcidicon[1]{\href{https://orcid.org/#1}{\mbox{\scalerel*{
\begin{tikzpicture}[yscale=-1,transform shape]
\pic{orcidlogo};
\end{tikzpicture}
}{|}}}}

\usepackage{xspace}            %

\begin{document}

\title{
A Comprehensive Tutorial on over 100 Years of Diagrammatic Representations of \\Logical Statements and Relational Queries 
}

\author{
\IEEEauthorblockN{Wolfgang Gatterbauer}
\IEEEauthorblockA{
\textit{Northeastern University}\\
Boston, USA \\
\orcidicon{0000-0002-9614-0504} 0000-0002-9614-0504}
}

\maketitle

\thispagestyle{plain}	%
\pagestyle{plain}		%

\begin{abstract}

Query formulation is increasingly performed by systems that need to guess a user's intent 
(e.g.\ via spoken word interfaces). 
But how can a user know that the computational agent is returning answers to the {``right''} query? 
More generally, 
given that relational queries can become pretty complicated, 
\emph{how can we help users understand relational queries}, whether human-generated or automatically generated?
Now seems the right moment to revisit a topic that predates the birth of the relational model:
developing visual metaphors that help users understand relational queries.

This lecture-style tutorial 
surveys the key \emph{visual metaphors developed 
for diagrammatic representations of logical statements and relational expressions},
across both the relational database and the much older diagrammatic reasoning communities.
We survey the history and state-of-the-art of relationally-complete 
diagrammatic representations of relational queries,
discuss the key visual metaphors developed in over a century of 
investigations into diagrammatic languages,
and organize the landscape by mapping the visual alphabets of diagrammatic representation systems to the syntax and semantics of Relational Algebra (RA) and Relational Calculus (RC).
Tutorial website:
\url{https://northeastern-datalab.github.io/diagrammatic-representation-tutorial/}
\end{abstract}

\section{Introduction}

The design of relational query languages and the difficulty for users to compose relational queries have
received much attention over the 
last 40 years~\cite{DBLP:journals/vlc/CatarciCLB97, 
ChanUserDatabaseInterface:1993,
Harel:Nonprocedural:1985,
FrameworkForChoosingQueryLanguages:1985,
LEGGETT1984493,
DBLP:journals/csur/Reisner81, 
Reisner1975:HumanFactors,
Welty-Stemple:1981,
scamell:1993}.
A complementary and much-less-studied problem is that of helping users 
\emph{read and understand existing queries}.  
With the proliferation of public data sources and associated queries, users increasingly have to read other people's queries and scripts.  
At the same time, Large Language Models (LLMs) have become an effective way to generate ``starter code'' (including SQL queries \cite{DBLP:journals/corr/abs-2308-15363:Text-to-SQL, DBLP:journals/corr/abs-2208-13629:ASurveyonText-to-SQLParsing}) which still has to be checked for correctness and undergo refinement.
Some even predict that
``\emph{all programs in the future will ultimately be written by AIs, with humans relegated to, at best, a supervisory role}''~\cite{Welsh:EndOfProgramming}.
But while it is easier to modify a ``starter query'' than to write something from scratch,
reading code is hard, and SQL is no exception.  
Any ``human-AI-database interaction'' relying on starter queries still requires users to understand written queries. 
For that reason alone, it is an opportune moment to study ways that help users understand queries,
and visualization is one obvious route.
While visual methods for composing queries have been studied
extensively in the database literature under the topic of Visual Query Languages (VQLs)~\cite{DBLP:journals/vlc/CatarciCLB97}, 
the challenges for supporting the \emph{reverse functionality} of automatically creating a visual representation of an existing query
(``Query Visualization'') are different than the problem of composing a new query.

\begin{figure}[tb]
\centering
\includegraphics[scale=0.5]{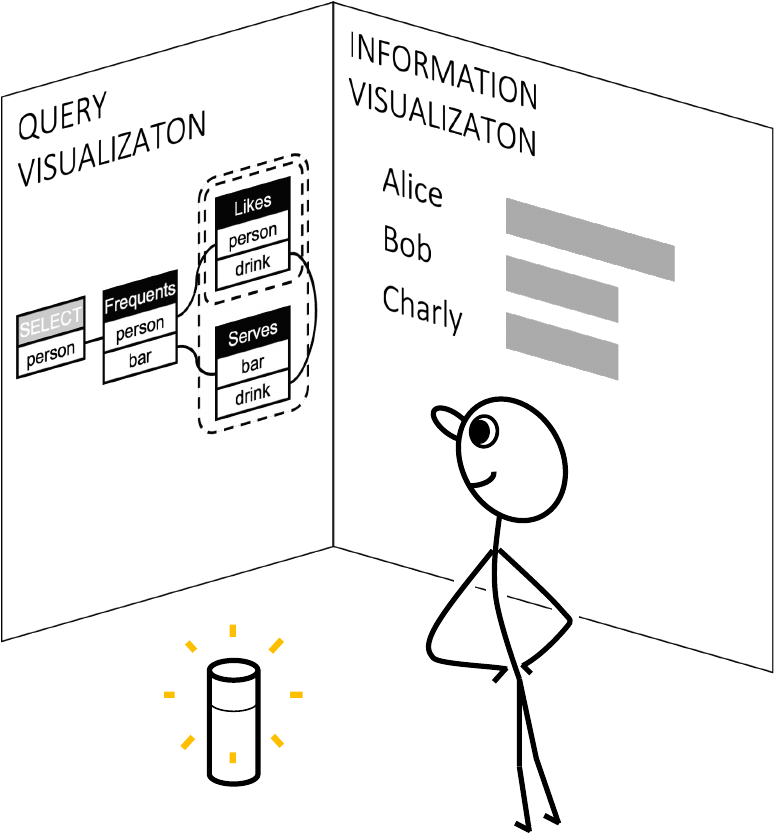}
\caption{An analyst dictates a query to her voice assistant which then shows the query as understood together with the query answers.}
\label{Fig_SpeechAssistantv2}
\end{figure}

\begin{figure}[tb]
\hspace{-3mm}
\includegraphics[scale=0.37]{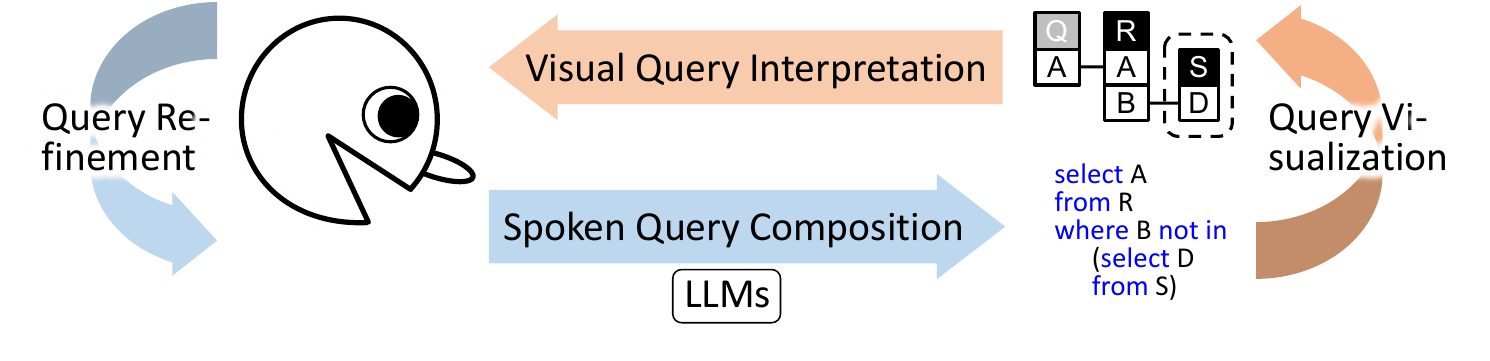}
\caption{The future user interaction with relational databases: a user dictates queries, 
and the system visualizes the queries back for the user to verify the correct interpretation.}
\label{Fig_New_Vision}
\end{figure}

The tutorial uses a few relational queries 
to survey and summarize the history of \emph{diagrammatic (thus visual) representations 
of first-order logic queries and statements}.
The goal is to highlight similarities and differences
between approaches proposed across communities by contrasting mapping of 
various visual representations to equivalent expressions in 
Relational Algebra (RA) and Relational Calculus (RC).

\smallsection{Outline}
The lecture-style 3-hour tutorial consists of six parts:

(1) \emph{Why visualizing queries and why now}:
	We contrast Query Visualization (QV) with Visual Query Languages (VQL) 
	and give several usage scenarios 
	for the use of query visualization
	(see~\cref{Fig_SpeechAssistantv2,Fig_New_Vision}).

(2) \emph{Principles of query visualization}:
	We discuss several proposed principles of query 	%
	visualization~\cite{GatterbauerDJR:PrinciplesOfQV:2022,gatterbauer2011databases},	
	re-phrased in the terminology of ``Algebraic Visualization Design''~\cite{DBLP:journals/tvcg/KindlmannS14}.
	We use these principles as guides when later discussing different diagrammatic representations.

(3) \emph{Logical foundations of relational query languages}:
	We discuss the logical foundations of relational query languages.
	We show a few example queries over the sailors-reserve-boats database
	in different relational query languages, which we later use when discussing and comparing
	visual representations.

(4) \emph{Early diagrammatic representations}:
	Diagrammatic representations for logical statements
	were developed well before relational databases. 
	We discuss several visual formalisms developed over more than 100 years of work on diagrammatic reasoning systems,
	in particular the influential beta existential graphs by Peirce~\cite{peirce:1933}
	and their connection to the much later developed Domain Relational Calculus (DRC).

(5) \emph{Modern visual query representations}:
	We use the earlier introduced queries over the sailors-reserve-boats database 
	to discuss the main families of visual representations for relational queries 
	proposed by the database community.

(6) \emph{Lessons learned and open challenges}: 
	We extract insights from our survey and discuss open challenges.

\section{Tutorial information}

\smallsection{Audience and prerequisite}
This 180~min tutorial targets
researchers and practitioners 
who desire an intuitive, yet comprehensive survey
of \emph{diagrammatic representations of logical statements and relational queries}.
Our focus is on the commonalities and differences between
major design ideas.
The tutorial is best followed by being familiar with Relational Algebra (RA), Relational Calculus (RC) and the safety conditions to make them equivalent in expressiveness.
However, the tutorial is self-contained and includes 
a short-paced summary of the characteristics of 5 relational query languages.

\smallsection{Scope of this tutorial}
This tutorial surveys visual formalisms for representing \emph{relational queries}. 
The focus is on relationally complete formalisms whose expressiveness is equivalent to Relational Algebra (RA),  Relational Calculus (RC), and non-recursive Datalog with stratified negation.
In order to guide the discussion, the tutorial discusses mapping the visual alphabets of visual formalisms 
to expressions of RA and RC. It thus starts with a quick fast-paced overview of RA and RC and their connections to first-order logic.

\smallsection{Out-of-scope}
The tutorial does not cover domain-specific visualizations, such as those for geographic information systems, time-series, and
spatio-temporal 
data~\cite{DBLP:conf/ieeevast/CorrellG16,DBLP:conf/chi/ManninoA18,DBLP:journals/tvcg/LeeLSKKP20}.
Neither does it cover 
dynamic
interaction with queries or data~\cite{10.14778/2732240.2732247}.

\smallsection{Related other tutorials}
A tutorial at SIGMOD'19~\cite{DBLP:conf/sigmod/TangWL19} (``Towards Democratizing Relational Data Visualizations'') focused on ways to visualize data and languages that allow users to specify what visualizations they want to apply to data. 
The focus of this tutorial is instead of visual representations \emph{of queries}.
Two tutorials at SIGMOD'17~\cite{DBLP:conf/sigmod/BhowmickCL17} (``Graph Querying Meets HCI'') and SIGMOD'22~\cite{DBLP:conf/sigmod/BhowmickC22} (``Data-driven Visual Query Interfaces for Graphs'') focused on visual composition of graph queries. 
The types of queries discussed in those tutorials basically correspond to conjunctive queries with inequalities over binary predicates, whereas our focus is on full first-order logic. 
Also, the focus was on the human-interaction aspect of how to compose queries, while our focus is on the visual formalisms developed for relational queries over the last century (thus even predating the relational model).

\smallsection{Contrast to prior offerings of this tutorial} 
A 90-min tutorial on the topic 
was presented at the ``{International Conference on the Theory and Application of Diagrams 2022}'' (DIAGRAMS-22)~\cite{gatterbauer:diagrams:tutorial:2022},
the main international venue covering all aspects of research on the theory and application of diagrams.
This conference attracts an audience with close to no intersection with the audience at database conferences. 
The emphasis of that tutorial was on the logical foundations of relational databases, 
the resulting different focus from the diagrammatic reasoning community,
and the problems arising from visualizing logical disjunctions as diagrams.
A 90-min version of the proposed tutorial 
was presented at VLDB 2023~\cite{DBLP:journals/pvldb/Gatterbauer23}.
The emphasis of that tutorial was on the principles guiding effective Query Visualization,
and an overview of representations suggested by the database community.
The tutorial is available as a 300-page slide deck on the tutorial web page~\cite{DBLP:journals/pvldb/Gatterbauer23}.

The key novel parts of this 3-h tutorial will be as follows:
\circled{1}~Part 4 includes a comprehensive survey of early diagrammatic approaches for representing logical statements
that largely predate attempts in the database community and are seemingly disconnected to them.
We will cover
Euler circles~\cite{Euler:1802},
Venn diagrams~\cite{venn:1880}, 
Venn-Peirce diagrams~\cite{peirce:1933},
constraint diagrams~\cite{kent:1997,GilHowseKent:1999},
Peirce's beta existential graphs, and
Sowa's conceptual graphs~\cite{Sowa:1976:CGD:1664383.1664387}.
We also discuss 
formalisms embodied by Higraphs~\cite{DBLP:journals/cacm/Harel88}
and UML notation~\cite{folwer:UMLdistilled:2003}.
\circled{2}~Part~5 compares past approaches in the database community. It was previously (at VLDB 2023) only covered partially due to the limited time. 
This tutorial includes an additional query
focusing on diagrammatic representations for disjunctions and unions, 
which are known to be the greatest challenge for diagrammatic representations (see e.g., discussion by Shin~\cite{shin_1995}). 
We will also cover two additional very recent diagrammatic formalisms.
Furthermore, the historical comparisons (parts 4 and 5 together) will culminate in a new ``lessons learned'' synthesis (part 6).
\circled{3} The new material grows the expected number of slides to over 400 pages. 
Slides (and possibly videos) of the tutorial will be made available afterward on the tutorial web page,
similar to other recent tutorials by the presenter and collaborators on unrelated topics~\cite{DBLP:conf/sigmod/TziavelisGR20,DBLP:conf/icde/TziavelisGR22}.

\section{Tutorial content}

\subsection{Part 1: Why visualizing queries and why now?}

We give several scenarios 
in which ``appropriate'' query visualizations could help users 
achieve new functionalities 
or increased efficiency in composing queries.
An important detail is here that visualizations can be used as \emph{complement} 
to query composition  
\emph{instead of substitution}.
This contrasts with 
Visual Query Languages (VQLs) which allow users to express queries in a visual format.
Visual methods for specifying relational queries have been studied
extensively~\cite{DBLP:journals/vlc/CatarciCLB97},
and
many commercial database products offer some visual interface for users to write simple conjunctive queries.  
In parallel, there is a centuries-old history of the study of formal diagrammatic reasoning systems \cite{DBLP:conf/iccs/Howse08}
with the goal of helping humans to reason in terms of logical statements.

Yet despite their  intuitive appeal and extensive study, successful visual tools today mostly only 
complement instead of replace
text for composing queries.
We will discuss several reasons why visual query composition
for general relational queries
have not yet widely replaced textual query composition
and discuss a user-query interaction 
that separates the query composition from the visualization:
Composition is either unchanged and still done in text, 
or alternatively replaced with natural language (NL) interfaces to personal assistants and learned models (\cref{Fig_SpeechAssistantv2,Fig_New_Vision}).
This composition is then augmented and \emph{complemented with a visual interaction that helps interpretation and verification of correctness}~\cite{gatterbauer2011databases}.

With this motivation, the goal of this tutorial is to survey and highlight the key ideas behind
major proposals for diagrammatic representations of relational statements and queries.

\begin{definition}[\textbf{Query Visualization}~\cite{GatterbauerDJR:PrinciplesOfQV:2022}]
The term query visualization refers to both ($i$) a graphical representation of a query 
(alternatively, ``\emph{query diagram}'')
and ($ii$) the process of transforming a query into a graphical representation
(alternatively, ``\emph{query diagramming}'').
The goal of query visualization is to help users more quickly understand the intent of a query,
as well as its relational query pattern.
\end{definition}

\subsection{Part 2: Principles of Query Visualization}

The challenge of query visualization is to find appropriate visual metaphors 
that ($i$) allow users to quickly understand a query's intent, even for complex queries,
($ii$) can be easily learned by users,
and ($iii$) can be obtained from textual queries by automatic translation, 
including a visually appealing automatic arrangement of elements of the visualization.
We discuss several earlier proposed principles of query visualization~\cite{GatterbauerDJR:PrinciplesOfQV:2022,gatterbauer2011databases},
which are newly organized, extended, and rephrased in the terminology of ``Algebraic Visualization Design''~\cite{DBLP:journals/tvcg/KindlmannS14}.
One important ``correspondence principle'' relies on a recently proposed notion of ``\textbf{relational query pattern}''~\cite{GD:RelationalDiagrams:2024}.
While we call them ``principles", they are not meant to be irrevocable axioms, but rather intuitive objectives, whose formulation helps us develop a shared vocabulary for later discussing the trade-offs among various visualizations.
We also include them in order to spark a healthy debate during and after the tutorial.

\subsection{Part 3: Logical foundations of relational query languages}

We give a 
brief
overview of the logical foundations of relational query languages
by discussing \emph{5 queries} over a variant of the sailors-reserve-boats database
from the ``cow book'' \cite{RamakrishnanGehrke:DBMS2000}.
We use a consistent notation and give the queries in \emph{5 textual query languages}: 
\textbf{SQL}, \textbf{Domain Relational Calculus (DRC)}, \textbf{Tuple Relational Calculus (TRC)}, non-recursive \textbf{Datalog} with negation, and \textbf{Relational Algebra (RA)}.
We use these queries and textual languages  later in parts 4 and 5
where we establish the mappings between various visual formalisms and these 5 queries. 
By using a consistent set of queries throughout our survey we can give a unified comparison of visual alphabets and their ``pattern expressiveness".
Our focus is on expressiveness equivalent to First-Order Logic (FOL), 
which allows us to make the connection to a century of research on formalisms for diagrammatic reasoning.

\subsection{Part 4: Early diagrammatic representations}

A query in Relational Calculus (RC) is a logical formula with free variables and as such
a specialization of First-Order Logic (FOL). 
A \emph{logical statement} has no free variables and is basically the same as a \emph{Boolean query} that returns a truth value of TRUE or FALSE. 
Diagrammatic representations for logical statements~\cite{DBLP:conf/iccs/Howse08} 
have been developed even before FOL, which was only clearly articulated in the 1928 first edition of David Hilbert and Wilhelm Ackermann's ``Grundzüge der theoretischen Logik''~\cite{HilbertAckerman:1928}.

An influential diagrammatic notation is the \textbf{Existential Graph (EG)} notation by Charles Sanders Peirce~\cite{peirce:1933,Roberts:1992,Shin:2002}, 
who wrote on graphical logic as early as 1882~\cite{Peirce:vo4:1879-1884}.
These graphs exploit topological properties, such as enclosure, to represent logical expressions and set-theoretic relationships.
Peirce's graphs come in two variants: 
alpha and beta.
Alpha graphs represent propositional logic, whereas beta graphs represent First-Order Logic (FOL). 
Both variants use so-called \emph{cuts} (simple closed curves) to express negation,
and beta graphs use a syntactical element called the \emph{Line of Identity} (LI) to denote 
\emph{both the existence of objects and the identity between objects}.
An important component of our discussions of beta-existential graphs is showing their
imperfect mapping to the Boolean fragment of Domain Relational Calculus (DRC).
As we show, this imperfection has led to a lot of follow-up and confusion in various works on Peirce's existential graphs.

We also cover
\textbf{Euler circles}~\cite{Euler:1802},
\textbf{Venn diagrams}~\cite{venn:1880}, and
\textbf{Venn-Peirce diagrams}~\cite{peirce:1933},
following mainly the exposition by Shin~\cite{shin_1995}.
We discuss
\textbf{constraint diagrams}~\cite{kent:1997,GilHowseKent:1999},
Sowa's \textbf{conceptual graphs}~\cite{Sowa:1976:CGD:1664383.1664387},
and formalisms embodied by \textbf{Higraphs}~\cite{DBLP:journals/cacm/Harel88}
and \textbf{UML notation}~\cite{folwer:UMLdistilled:2003}.
We may or may not cover \textbf{Frege's two-dimensional conceptual notation}~\cite{Frege:1879}.

\subsection{Part 4: Modern Visual Query Representations}

We discuss the main proposed visual representations for relational queries. 
We will also include influential Visual Query Languages (VQLs) as long as those support (either directly or via simple additions) the inverse functionality of visualizing an existing relational query.
A key difference of our tutorial in contrast to all prior surveys and overviews that we are aware of (like \cite{DBLP:journals/vlc/CatarciCLB97}) is 
that this tutorial shows original figures by using 
a consistent schema (the sailor-boat-database from the ``cow book'' \cite{RamakrishnanGehrke:DBMS2000}) 
and a few intuitive queries (such as ``find sailors who have rented all red boats'') 
to provide a consistent comparison across different past proposals.

\textbf{Query-By-Example (QBE)}~\cite{DBLP:journals/ibmsj/Zloof77} is an influential early VQL that was influenced by DRC.
QBE can express relational division by
breaking the query into two logical steps and using a temporary relation~\cite[Ch. 6.9]{RamakrishnanGehrke:DBMS2000}.
In doing so, QBE
uses a query pattern from 
Datalog of implementing relational division (or universal quantification)
in a dataflow-type, sequential manner, 
requiring multiple occurrences of the same table.
We compare queries in QBE against equivalent Datalog queries and ask whether QBE is really more ``visual'' than Datalog.

\textbf{Interactive query builders} employ 
visual diagrams that 
users can manipulate (most often in order to select tables and attributes)
while using \emph{a separate query configurator}
(similar to QBE's condition boxes~\cite{DBLP:journals/ibmsj/Zloof77}) 
to specify selection predicates, attributes, and sometimes nesting between queries.
They work mainly for constructing conjunctive queries
but limited forms of negation and union can be incorporated into the condition part of such queries. 
For more general forms of negation and union, however, views as intermediate relations need to be used, resulting in multiple screens.
dbForge~\cite{dbforge} is the most advanced and commercially supported tool we found for interactive query building.
Yet it does not have a visual formalism for non-equi joins between tables 
and the actual filtering values and aggregation functions can only be added in a separate query configurator.
Moreover, it has limited support for nested queries: 
the inner and outer queries are built separately,
and the diagram for the inner query is \emph{presented separately and disjointly} 
from the diagram for the outer query.
Thus  \emph{no visual depiction of correlated subqueries is possible}.
Other graphical SQL editors such as SQL Server Management Studio (SSMS)~\cite{ssms}, Active Query Builder~\cite{activequerybuilder}, QueryScope from SQLdep~\cite{queryscope}, MS Access \cite{msAccess}, and
PostgreSQL's pgAdmin3~\cite{pgadmin} lacks in even more aspects of visual query representations: 
most do not allow nested queries, 
none has a single visual element for the logical quantifiers 
\texttt{NOT} \texttt{EXISTS} or \texttt{FOR} \texttt{ALL},
and all require specifying details of the query in SQL or across several tabbed views 
\emph{separate from a visual diagram}.

\textbf{Dataflow Query Language (DFQL)} 
is an example visual representation that is relationally complete \cite{DBLP:journals/iam/ClarkW94,DBLP:journals/vlc/CatarciCLB97}
by mapping its visual symbols to the operators of relational algebra.
Following the same procedurality as RA, DFQL expresses the data flow in a top-down tree-like structure. 
Like most visual formalisms
that we are aware of and that were proven to be relationally complete
(including those listed in
\cite{DBLP:journals/vlc/CatarciCLB97})
they are at their core visualizations of relational algebra operators.

\textbf{Query By Diagram (QBD)}~\cite{DBLP:journals/tse/AngelaccioCS90,DBLP:journals/vlc/AngelaccioCS90, DBLP:conf/sigmod/CatarciS94}
is based on an ER (Entity-Relationship) model of the data.
\textbf{TableTalk}~\cite{DBLP:journals/vlc/Epstein91}
visualizes the flow of a query top-down and 
displays logical conditions in tiles.
\textbf{Object-Oriented VQL}~\cite{DBLP:journals/tkde/MohanK93}
adds existential and universal quantifiers to attributes.

\textbf{Visual SQL}~\cite{DBLP:conf/er/JaakkolaT03} is 
a visual query language that also supports query visualization. 
With its focus on query specification, it maintains the one-to-one correspondence to SQL,
and syntactic variants of the same query lead to different representations.
Similarly, \textbf{SQLVis}~\cite{DBLP:conf/vl/MiedemaF21} 
places a strong focus on the actual syntax of SQL queries, 
and syntactic variants like nested EXISTS change the visualization.

\textbf{QueryVis} (earlier \emph{QueryViz})~\cite{DanaparamitaG2011:QueryViz, 
DBLP:journals/tvcg/BartolomeoRGD22,
gatterbauer2011databases,
DBLP:conf/sigmod/LeventidisZDGJR20}
borrows the idea of a ``default reading order''
from diagrammatic reasoning systems~\cite{DBLP:conf/diagrams/FishH04} 
and uses \emph{arrows} to indicate an implicit reading order between different nesting levels.
Without the arrows, there would be no natural order placed on the existential quantifiers 
and the visualization would be ambiguous.

\textbf{DataPlay}~\cite{DBLP:conf/uist/AbouziedHS12} 
uses a nested universal relation data model and
allows a user to compose their query by interactively modifying a \emph{query tree with quantifiers}
and observing changes in the matching/non-matching data.

\textbf{SIEUFERD}~\cite{DBLP:conf/sigmod/BakkeK16}~is a direct manipulation spreadsheet-like interface that lets users manipulate the actual data.
In SIEUFERD, a result header encodes ``the structure'' of the query.
The query result is listed below that header.

\textbf{String diagrams}~\cite{DBLP:conf/diagrams/Haydon2020:StringDiagrams,StringDiagrams:2024:arxiv}
are essentially a variant of Peirce's beta graphs 
that allow free variables in addition to bound variables. 
Both types of variables are represented by lines, yet bound ``variable lines'' end in a dot.

\textbf{Relational Diagrams}~\cite{GD:RelationalDiagrams:2024}
are a recent variant inspired by QueryVis that indicates the nesting structure of table variables by using \emph{nested negated bounding boxes} (instead of arrows)
inspired by Peirce's beta existential graphs.
Interestingly, because \emph{Relational Diagrams} are based on Tuple Relational Calculus (instead of Domain Relational Calculus which is closer to First-Order Logic)
they solve interpretation problems of Peirce's beta graphs that have been the focus of intense research in the diagrammatic reasoning community.

\subsection{Part 6: Lessons Learned and Open Challenges}
By extracting and synthesizing insights from our survey, 
we give design choices that avoid ambiguities resulting from overloading the meaning of lines as geometric marks (dubbed ``the 3 abuses of the line"),
and discuss open challenges.

\section{Author information}

\textbf{Wolfgang Gatterbauer} is 
an Associate Professor at the Khoury College of Computer Sciences at Northeastern University. His research interests lie in the intersection of theory and practice of data management. He received an NSF Career award and -- with his students and collaborators -- a best paper award at EDBT 2021, best-of-conference mentions for PODS 2021, SIGMOD 2017, WALCOM 2017, and VLDB 2015, and two reproducibility awards for papers from SIGMOD 2020.

\subsection{Acknowledgements}

This work was supported in part by the 
NSF
under award numbers CAREER IIS-1762268 and IIS-1956096.

\bibliographystyle{IEEEtranS} 
\bibliography{queryvis-icde-tutorial.bib}

\end{document}